# Toward a Science of Autonomy for Physical Systems: Construction

Miroslaw Skibniewski
mirek@umd.edu
University of Maryland, College Park

Mani Golparvar-Fard
mgolpar@illinois.edu
University of Illinois at Urbana-Champaign

## Introduction

Today, ensuring and improving safety, productivity, quality, and sustainability in construction, operation, and maintenance of national civil infrastructure systems through advances in robotics and automation is a national imperative [1,2]. By "national civil infrastructure" we refer to the 4.5M commercial buildings, 3.9M miles of public roads, 2M miles of oil and natural gas pipelines, 600K bridges, 190K cell phone towers, 120K miles of major railroads, 100K miles of levees, 84K dams, 50K miles of electrical power lines, 25K miles of commercially navigable waterways, and 5K public-use airports in the United States, all of which are critical to our national economy and society. By "ensuring and improving safety, productivity, quality, and sustainability" we mean to guarantee performance across the entire life cycle of civil infrastructure, from construction to operation and from maintenance to disposal. By "advances in robotics and automation" we mean innovative research and education that will provide tools for systematic and timely execution of construction, and also collection, analysis, visualization, and operational use of big data in built environments for project monitoring and control purposes. In the following, we discuss the current state of construction and operation of the national civil infrastructure systems in detail and present several opportunities for improvements through research and education on robotics and automation.

## Construction industry

Construction is a $950 billion industry [3], with 25-50% waste in coordinating labor and equipment, and in managing, moving, and installing material [4-9], and a productivity which has been declining for more than 30 years [4,10-12]. This is particularly troubling when we consider that the quality of life of every American relies in part on the output of the construction industry: residential, commercial and instructional buildings, airports, refineries, roads, bridges, power plants, water and sewer lines, and other infrastructure [1].

---

To insure our nation's social and economic prosperity, in 2008 the National Institute of Standards and Technology (NIST) appointed an ad-hoc committee of experts through the National Research Council (NRC) of the National Academies to provide advice for advancing the competitiveness and productivity of the U.S. construction industry. Among the recommendations provided by this committee, two groups stand out:

*a) Greater use of robotics and automation in execution of construction projects:*
Automated equipment for excavation and earthmoving operations, concrete placement, and pipe installation could significantly cut waste, improve jobsite safety and the overall quality of construction projects. Among these, improving safety is most imperative as the construction industry exhibits daily average of four accidents on every site in the US (a total of 4,045 deaths in 2013 alone). Operating construction equipment represent one of the most dangerous tasks on construction sites. Often accidents related to their operations result in death and catastrophic destruction on the construction site and the surrounding area. Such accidents also cause considerable delays and disruption, and thus negatively impact the efficiency of the operations. Among commonly used equipment, cranes have a larger share in accidents. For instance, according to the U.S. Bureau of Labor Statistics census of fatal occupational injuries, between 1992 and 2006 there were 632 crane-related deaths from 610 construction incidents, averaging 43 deaths per year. In 2008 alone, there were 54 worker and 4 bystander deaths and 100 construction worker, 15 bystander and 11 first responder injuries related to construction crane accidents.

To improve safety and productivity in construction equipment, concepts of, and experiments with, semi-autonomous single- and multiple-purpose construction robots have been developed since 1980's [13-19]. However, practical implementation to date has been difficult due to the fragmentation of construction processes and high cost of equipment mobilization. New construction site data acquisition and processing technologies and widespread application of Building Information Modeling (BIM) provide a good prospect for overcoming these limitations.

b) *Effective performance monitoring to drive efficiency and support innovation:*
Measuring performance through the execution of a project is an enabler of innovation. Quick and easy detection, assessment, and communication of performance deviations provides project management with an opportunity to take corrective actions that can minimize negative impacts on budget, schedule, quality, and safety of a project. More specifically, cost and delivery time in construction projects can be significantly reduced with tools that better characterize the extent to which construction plans are being followed (*progress monitoring*) and the extent to which workers and equipment are being fully utilized (*activity analysis*). Implementing these performance monitoring schemes in a systematic manner helps companies and organizations understand how processes or practices led to improvements or inefficiencies. it also enables them to use that knowledge to improve products, processes, and the outcomes in the upcoming operations or other active projects.

To implement performance measurements, construction companies already collect data through various sensing modalities and assign field engineers to filter, annotate, organize, and present the data. However, the cost and complexity of the collection, analysis and reporting operations result in sparse and infrequent monitoring, and thus some of the gains in efficiency are consumed by monitoring costs. Individual construction companies do not have the expertise to improve automation or the incentive to invest in the research needed to reduce construction costs nationally. A multi-layer commitment to creating and developing new methods for sensing, analytics, and visualization of construction performance metrics is needed, encompassing technical and financial support at the federal level, state level, and at the level of construction companies.

## Autonomy in Condition Assessment

Maintaining and operating national civil infrastructure systems also faces several critical challenges. For example, of the 600K bridges in the United States, almost 70K (more than 11%) are classified as "structurally deficient" by the Federal Highway Administration (FHWA). In the state of Illinois alone more than 8M cars travel over deficient bridges every day, risking disaster similar to the I-35W Mississippi River bridge collapse in Minnesota in 2007.

More than $70B is needed to fix these bridges, and this figure is likely to rise as "many of our most heavily traveled bridges—including those built more than forty years ago as part of the interstate system—near the end of their expected lifespan." National budget priorities do not allow this level of funding, so many bridges will be left deficient. Deciding which ones to fix (and when) requires bridge inspection, a process that is currently done by human workers, that involves various safety hazards, and that is costly and prone to error. New site data acquisition methods such as Unmanned Aerial Vehicles (UAVs) equipped with cameras and laser scan and processing technologies such as methods for generating point cloud models, producing semantically rich CAD modeling from point cloud data, and automatically detecting, classifying, and localizing defects, and mapping them to various forms of condition index provide a good prospect for overcoming these limitations.

## The Path Forward in Research and Education

The Architecture/Engineering/Construction and Facility Management (AEC/FM) industry is still far from having agents and systems that provide the breadth and depth of the required capabilities mentioned above. At the fundamental level, planning and designing resilient systems that can account for all performance variabilities such as quality, safety, efficiency, and sustainability is difficult. Also creating sensing and analytics that can capture, synthesize, and represent actual performance metrics, map them to expected performance at both system-level and at agent-level, and reason about performance deviations is still an emerging science. Considering them together as the elements of a cyber-physical system and within a project controls framework, a number of research areas/needs will be structured. The fundamental examination of these areas can address the current technical and socio-technical challenges in the AEC/FM industry and minimizes the risk of introducing autonomy [20,21]:

a. **Performance Sensing and Analytics:** More advancement is needed in the areas of sensing, pre-processing, analysis and fusion of actual performance data, intelligent searching and information retrieval, parallel and distributed computing, and knowledge management and discovery. Particularly techniques are needed to accurately and comprehensively capture and synthesize real-world (e.g. through images, videos, laser scans, speech signals, textual information, GPS, etc.) and convert it to the information and knowledge that can capture performance at both levels of civil infrastructure systems, and the agents that are involved in constructing and operating them.

b. **Performance Information Modeling and Simulation:** To manage the risk associated with automation and application of autonomous systems, more research is needed in      (a) *simulation* to understand, model, and reason about individual operations during construction and operation phases of a civil infrastructure systems (e.g. earthmoving);   (b)  *distributed simulation platforms* that couple independent simulation models to improve reliability of performance across complex inter-connected systems; and          (c) *modeling the hierarchy of multimodal performance data sources and relevant information* for simulation, distributed simulation, and model representations. A more advanced state of science in these fronts can ultimately facilitate storage and querying of real time performance data for planning, monitoring, and controlling of both civil infrastructure systems and the agents that are involved in their construction and operation.

c. **Autonomous Operations:** Addressing long-standing practical limitations associated with the application of autonomous systems (e.g. earthmoving equipment) in construction and operation of the civil infrastructure systems can ensure and improve their safety, productivity, quality, and sustainability. With the recent advent in site data acquisition and processing technologies and widespread application of BIM, the timing is perfect for fundamental research on (a) *engineering of autonomy* to make these systems more predictable, reliable, and trustworthy; (b) *interactions between autonomous systems and human agents* (both those who control them, and those who interact with them as co-workers).

*For citation use*: Skibniewski M. & Golparvar-Fard M. (2015). *Toward a Science of Autonomy for Physical Systems: Construction*: A white paper prepared for the Computing Community Consortium committee of the Computing Research Association. http://cra.org/ccc/resources/ccc-led-whitepapers/

*This material is based upon work supported by the National Science Foundation under Grant No. (1136993). Any opinions, findings, and conclusions or recommendations expressed in this material are those of the author(s) and do not necessarily reflect the views of the National Science Foundation.*